\documentclass[12pt]{iopart}

\usepackage{iopams}

\usepackage {graphicx}

\usepackage {csquotes}

\usepackage{color}

\usepackage{soul}

\begin{document}

\title {Pilot system development in metre-scale laboratory discharge}

\author{Pavlo Kochkin$^1$, Nikolai Lehtinen$^1$, Alexander (Lex) P.J. van Deursen$^2$, Nikolai {\O{}}stgaard$^1$}

\address{$^1$ Birkeland Centre for Space Science, University of Bergen, Bergen, Norway}
\address{$^2$ Electrical Engineering Department, Eindhoven University of Technology, Eindhoven, The Netherlands}

\ead{pavlo.kochkin@uib.no, Nikolai.Lehtinen@uib.no}

\begin{abstract}

The pilot system development in metre-scale negative laboratory discharges is studied with ns-fast photography. The systems appear as bipolar structures in the vicinity of the negative high-voltage electrode. They appear as a result of a single negative streamer propagation and determine further discharge development. Such systems possess features like glowing beads, bipolarity, different brightness of the top and bottom parts, and mutual reconnection. A 1D model of the ionization evolution in the spark gap is proposed. In the process of the nonlinear development of ionization growth, the model shows features similar to those observed. The visual similarities between high-altitude sprites and laboratory pilots are striking and may indicate that they are two manifestations of the same natural phenomenon.

\end{abstract}

\vspace{3 cm}

Version of \today \\

\vspace{2 cm}
{\color{red}
\large
Please cite this paper \\ http://iopscience.iop.org/article/10.1088/0022-3727/49/42/425203}

\maketitle

\section{Introduction}

Pilot systems play an important role in negative discharge development process. In metre-scale spark pilots appear as bipolar formations from a point in space called ``space stem'' and grow in both directions away from and towards the high-voltage electrode. In longer gaps pilots can transform into space leaders \cite{Gallimberti2002}. Such space leaders also appear in front of a lightning leader channel, grow in both directions and finally attach to the main channel. The stepped propagation is a common feature of both negative long laboratory sparks and lightning leaders. Lightning leader steps are also closely associated with discrete intense bursts of X-ray radiation \cite{Dwyer2005b} and may even be responsible for Terrestrial Gamma-Ray Flashes (TGFs), as was previously suggested in \cite{Cummer2005,Cummer2011,Stanley2006,Østgaard2013,Shao2010,Lu2010} and recently modeled in \cite{Celestin2015}. Similarly, as shown in \cite{Kochkin2015}, pilots are involved in X-ray burst generation in long laboratory sparks. Thus, experimental study of pilot system formation and development can provide more information about lightning leader X-rays and TGFs. However, our understanding of pilots is mostly based on streak photographs obtained in the last century.

The existence of bipolar structures in long laboratory discharges was first shown in 1960's \cite{Stekolnikov1962}. In 1981 the Les Renardi\^{e}res group performed a fundamental study on negative discharges using various electrodes, gap distances and voltage rise times \cite{LesRenardieres1981}. This study provided the first systematic description of the various phases of negative discharge development. Phenomena such as negative leader, space stem, pilot system, and space leader were photographed, identified and described.

Theoretical efforts and models to explain long atmospheric discharges were presented by Gallimberti (\cite{Gallimberti2002} and citations therein) but these do not explain how pilot systems form and develop \cite{Bazelyan2000}. It is assumed that pilots appear from a ``space stem'' ahead of the leader tip in virgin air. In 2003 Vernon Cooray \cite{Cooray2003} formulated the situation as follows: \textit{``The pilot system consists of a bright spot called the space stem of short duration, from which streamers of both polarity develop in opposite directions''}. This is consistent with our observations. We will demonstrate below that pilots are preceded by negative streamer heads and often contain several bright spots.\\

In this work we first show the pilot system development in the laboratory with high spatial and temporal resolution from the very beginning till attachment to the HV electrode. It is demonstrated how a single negative streamer creates such a complex bipolar structure. Then we propose a 1D model of the collective ionization front evolution that is capable to capture most important observed details, such as glowing beads, bipolarity and ionization front collisions. Striking similarity between pilots and high-altitude sprites will be discussed at the end.

\section{Experimental setup}

The setup was available at the High-Voltage Laboratory in Eindhoven University of Technology from 2008 till 2014 but it is currently dismantled. A 2.4~MV Haefely Marx generator was used to create metre-long sparks. The voltage was set at 1~MV with 1.2/50~$\mu$s rise/fall time when not loaded. The generator was connected to a spark gap between two conical electrodes. The setup and all measuring equipment was exhaustively described in series of publications \cite{Nguyen2008,Kochkin2012,Kochkin2014,Kochkin2015} and only its optical part will be repeated here for consistency.

To obtain images of the pre-breakdown phenomena between two electrodes, a ns-fast 4~Picos camera \cite{StanfordComputerOptics} was located at 4~m distance from the gap, perpendicular to developing discharge. The camera contains a charge coupled image sensor preceded by a fast switched image intensifier (ICCD). The image intensifier is a micro-channel plate that allows adjustment of the camera sensitivity by varying the applied voltage between 600 and 1000~V. The CCD is read out with 12-bit and 780x580 pixels resolution. Lenses were either Nikon 35~mm F2.8 fixed focus or Sigma 70-300~mm F/4–5.6 zoom. The field of view of the camera covers the region below the HV electrode. The camera has a black and white CCD and is not calibrated. The applied color scheme is linked to the light intensity and intended to increase visual perception, but it does not represent the actual plasma temperature. The camera was placed inside an EMC cabinet. Appropriate shielding protected the camera and its communication cables against electromagnetic interference. More EMC aspects of the setup have been discussed in \cite{VanDeursen2015}.

\section{The pilot system and its features}

\subsection{Pilots and X-ray bursts}

In 2009 Cooray et al. proposed a mechanism of X-ray generation in long laboratory sparks \cite{Cooray2009}. It was suggested that the X-ray bursts are caused by encounters of negative and positive streamer fronts. Although the negative discharge development process was simplified in this model, the main idea of streamer encounter as the emission source has recently been experimentally supported. With positive high-voltage pulse, X-rays indeed appear at the moment when positive corona from the HV electrode merge with negative corona from the grounded electrode \cite{Kochkin2012}. Many encounters between individual streamers of opposite polarity occur at this moment. The development of negative discharges is more complex. We photographed the X-ray source region with ns-fast camera, and showed that positive streamers appear in the vicinity of the negative HV electrode \cite{Kochkin2014}. The positive streamers originate from bipolar pilot systems and encounter nearby negative streamers. For more details and properties of the X-ray emission we refer to \cite{Kochkin2015,March2010,March2011}; a statistical analysis is given in \cite{Carlson2015}. It is assumed that the X-rays are generated by high energy electrons in Bremsstrahlung process. The first attempt to measure such electrons directly has recently been published in \cite{Østgaard2016}.

Here it should be noted that a recent simulation of the X-ray emission questioned the streamer encounter mechanism \cite{Ihaddadene2015}. While it was confirmed that such encounters dramatically increase the electric field between two streamer fronts to values much higher than the cold runaway breakdown, the field persists only for several picoseconds due to rapid rise of electron density. The ionization quickly collapses the field, giving no time for the electrons to accumulate high energy. It is possible, however, that the action of electric field on electron acceleration was underestimated in that work, as discussed in subsection~\ref{ssec:discuss_xray}. Nevertheless, in measurements the X-rays bursts and pilots coincide in space and time, so the precise role of the latter requires further investigation.

\begin{figure*}[ht]
\centering
\includegraphics[width=\linewidth]{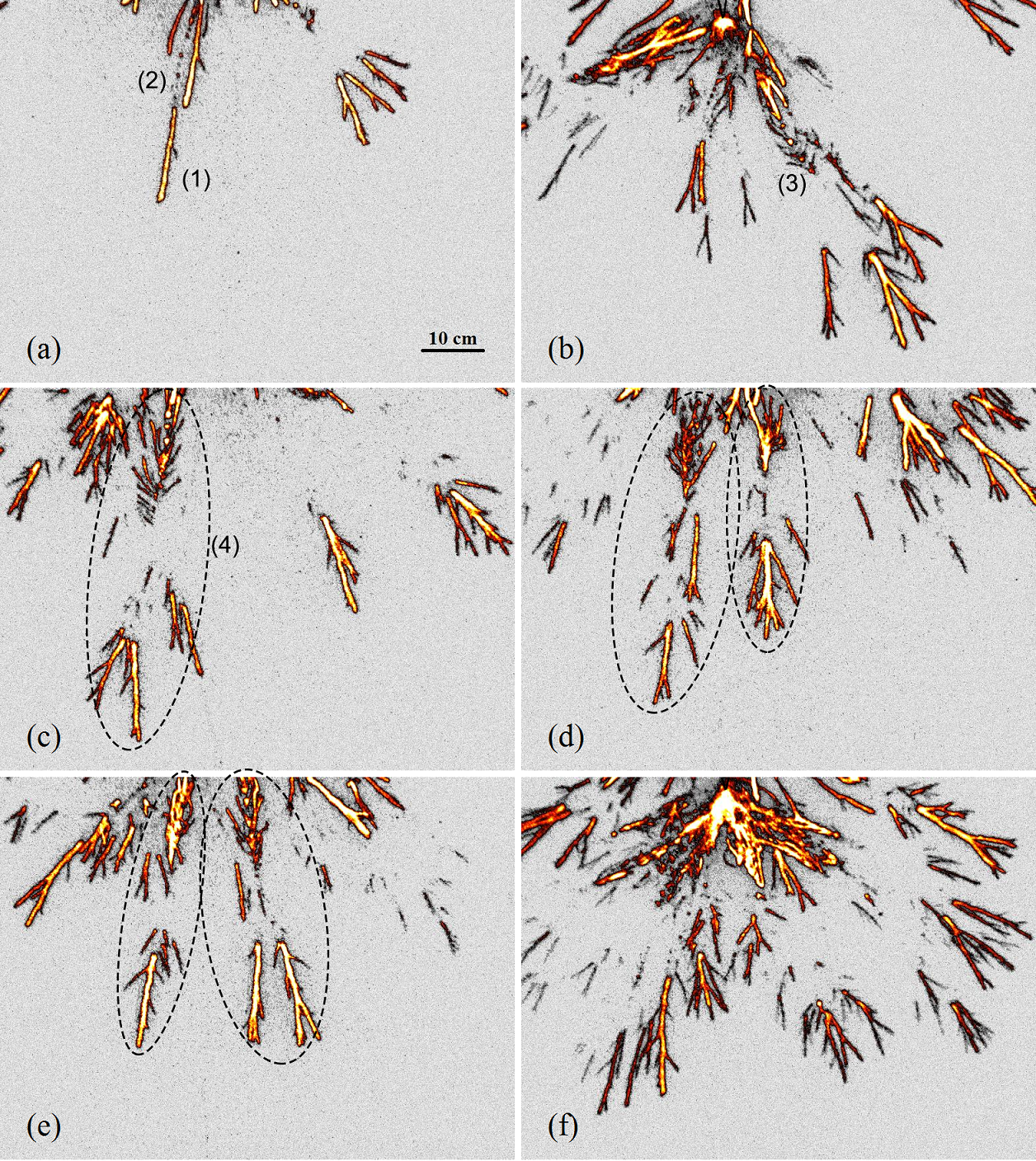}
\caption{The pilot system development process. The shown area is located below the HV electrode tip. The tip is only visible in images (b) and (f). In all images: (1) - a negative streamer (2) - beads (3) - a positive streamers (4) - a pilot system. All images have been exposed for 50~ns and show six different discharges at the moment of the most intense X-ray emission. }
\label{fig:collage}
\end{figure*}

\subsection{Pilot development}

\begin{figure*}[ht]
\centering
\includegraphics[width=\linewidth]{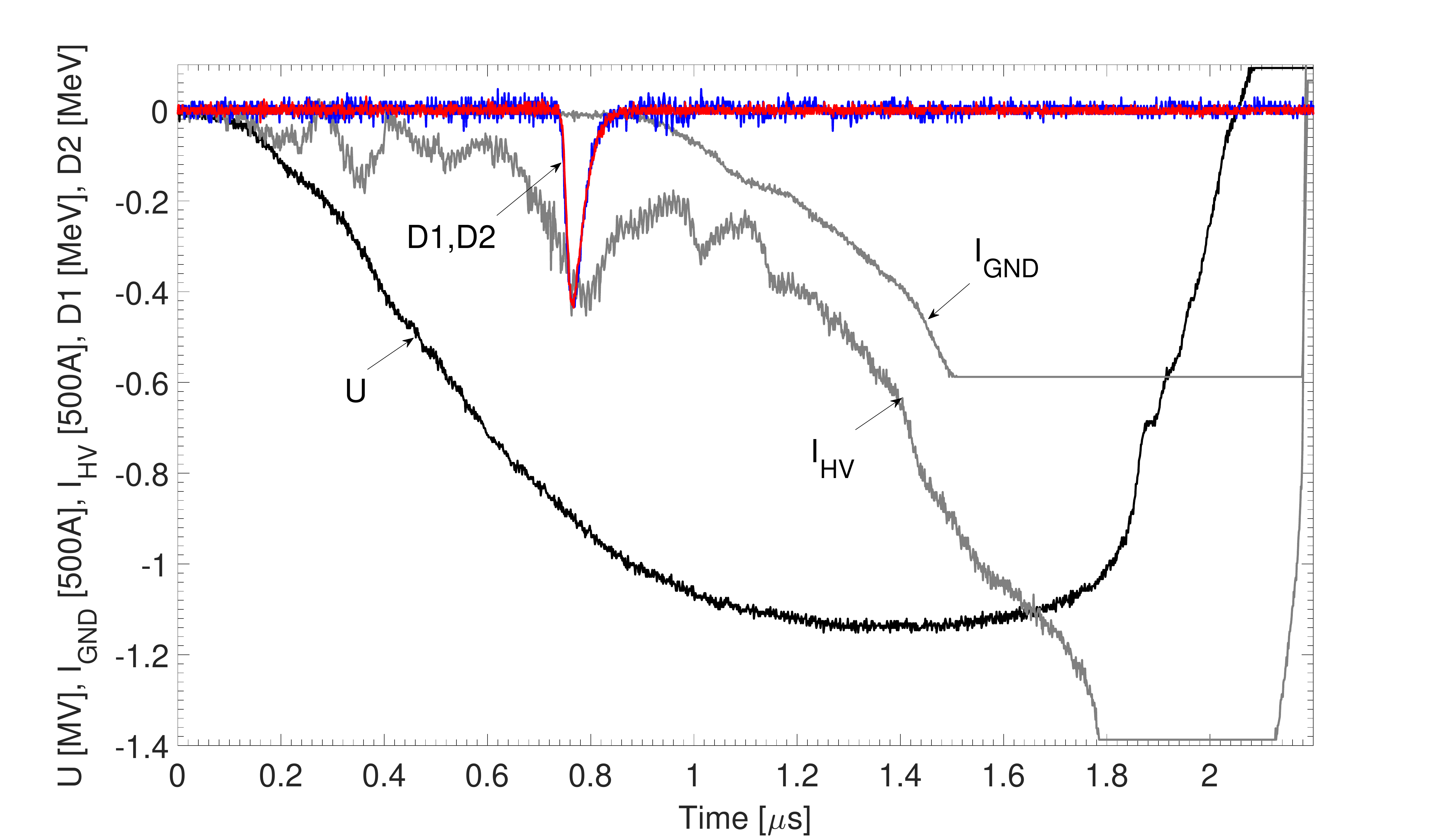}
\caption{The electrical characteristics and X-ray detection from the negative discharge. Voltage starts rising at $t=0~\mu$s, reaches its maximum of 1.1 MV at $t=1.4~\mu$s, and collapses at $t=1.8~\mu$s. The high-voltage current starts rising immediately, while the current through the grounded electrode appears later at $t=0.9~\mu$s. Two LaBr3 X-ray detectors show signals simultaneously and of the same 400~keV energy. }
\label{fig:plot}
\end{figure*}

For the sake of consistency, we reproduce here the pilot development as reported in \cite{Kochkin2014,Kochkin2015}, and discuss additional features. Figure~\ref{fig:collage} shows the pilot system development process. Every image was exposed for 50~ns and represents a single individual discharge at the moment of the most intense X-ray emission. The time delay between two consecutive discharges was at least 10~s. The depicted area is located below the HV electrode. The HV electrode tip is only visible in images (b) and (f) at the top in the middle.
The electrical signals of the full discharge are shown in figure~\ref{fig:plot}, the voltage $U$ over the gap, the currents $I_{HV}$ and $I_{GND}$ through both electrodes and the x-ray signals. Two LaBr$_3$ scintillation X-ray detectors were placed next to each other. Both simultaneously registered a 400~keV signal. All images of figure~\ref{fig:collage} are taken with 50~ns shutter time that fell within the x-ray time window, or between $t=0.7 - 0.8~\mu$s. The pilot systems occur at an advanced stage of the discharge development.
Returning to figure~\ref{fig:collage}, we observe the negative streamers (1) that originate from the HV electrode (image (a)). Some of them leave isolated beads (2) behind during the propagation. We will call them ``streamer beads'' or just beads; the reader should not confuse these with bead lightning \cite{Barry1980}.
In 2D images the beads appear at $2.5\pm0.4$~cm intervals. Some beads become branching points of the negative streamer. Eventually the channels and heads of the branched streamers form the negative streamer corona.

The streamer propagation velocity is measured by two different techniques: (i) by measuring the streamer trace length in an image with known exposure time and (ii) by measuring the displacement of a streamer head comparing two consecutive short exposure images taken by two cameras. The velocities are measured for many different streamers in different discharges; we further refer to \cite{Kochkin2014} for details on negative streamer and corona development. We found that the negative streamer heads propagate at $(3.7\pm0.2)\times$10$^6$~m/s, which is in agreement with previously reported in \cite{LesRenardieres1981} for 2~m gaps. Since the projection into the camera plane always reduces distances, the actual distance and velocity is likely to be closer to the highest measured value than to the average.
At the same moment, positive streamers start at the beads, either as single intense streak or as branches. The branches first move perpendicularly (image (c)) to the streamer channel driven by the streamer electric field, and then they start to propagate towards the HV electrode (images (b) - (f)). This gives the branches the appearance of a stack of the Greek letters $\Psi$. To substantiate the movement towards the electrode, we used two high-speed cameras triggered in sequence; the results have earlier been published in \cite{Kochkin2014} and are reproduced here for consistency.

\begin{figure*}[ht]
\centering
\includegraphics[width=0.5\linewidth]{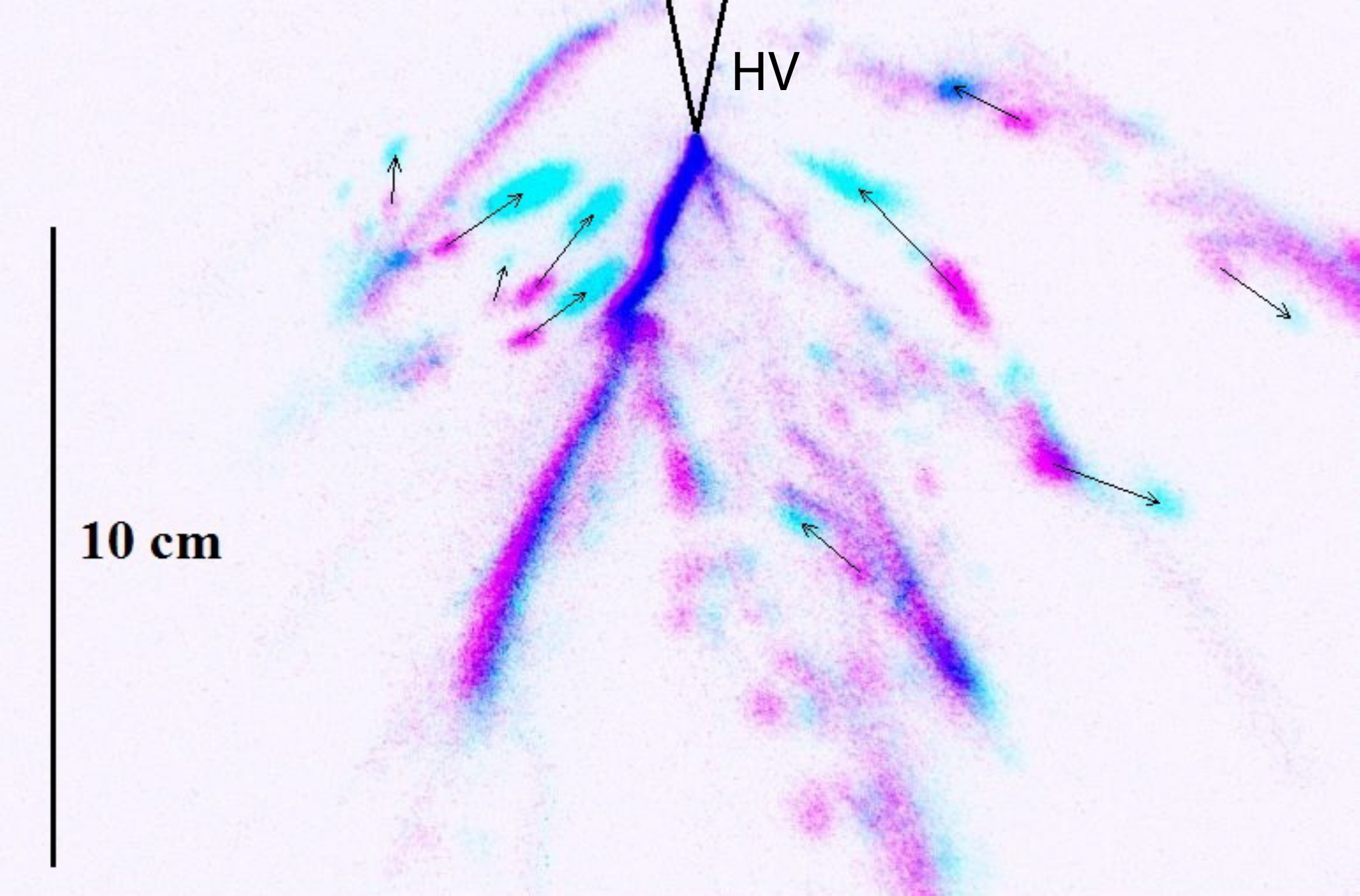}
\caption{Two images combined into one RGB picture. The first image is placed on the red layer, the second one on the blue layer. The exposure time is 3~ns for each image. The delay between two images is 10~ns. Arrows indicate the direction of motion. Figure reproduced from \cite{Kochkin2014}.}
\label{fig:two_cameras}
\end{figure*}

Figure~\ref{fig:two_cameras} shows two subsequent images made with 3~ns exposure time. The first image was placed on the red layer of an RGB picture, the second image was delayed by 10~ns with respect to the first one, and placed on the blue layer of the same picture. The arrows indicate the displacement from the red to the blue images. Clearly, many streamers move towards the cathode. The positive streamer velocity is difficult to measure correctly due to the limited extension in space and apparent dependence on other factors, such as the proximity of the HV electrode. We estimated velocity of the positive streamer head at $(1.5\pm0.7)\times$10$^6$~m/s, or half the speed of the negative ones. As a result, positive streamers appear brighter in the images than negative --- assuming equal intrinsic brightness. The entire structure - the negative corona, streamer beads with $\Psi$s, is called a 'pilot system' (4) in \cite{Gallimberti2002}. Pilots are encircled in images (c) - (e) by dashed ellipses.

In our setup the last bead appears at $36.8\pm6.3$~cm distance from the tip. We can use the streamer velocity data of figure~6 in \cite{Kochkin2014} to go back to the moment that the negative streamer head was at this point (0.5~$\mu$s in figure~\ref{fig:plot}). Then the applied voltage $U$ was 500~kV, or the local electric field $E\thickapprox12$~kV/cm assuming a $1/r$-dependence for $E/U$. This field is close to the  so-called ``stability field'' $E_s$ \cite{Babaeva1997}. It also indicates that the streamer experiences a shortage in charge and electric field which hinders smooth propagation. The blockade may lead to beading. As a support for this suggestion we refer to \cite{LesRenardieres1981} where it was demonstrated that a shorter voltage rise time, i.e. larger $dE/dt$, leads to smoother discharge development.

The upper parts of figure~\ref{fig:collage}(c) and (d) show many $\Psi$s that are about to collide with negative streamers or the cathode. Such collisions provide the kick-off for the few electrons that become run-away in the electric field and produce X-ray by Bremsstrahlung a few nanosecond later \cite{Kochkin2016}. We observed high frequency cathode current oscillations \cite{Kochkin2015} simultaneously with the x-rays, and also attribute the oscillation to the collisions.

\begin{figure*}[ht]
\centering
\includegraphics[width=\linewidth]{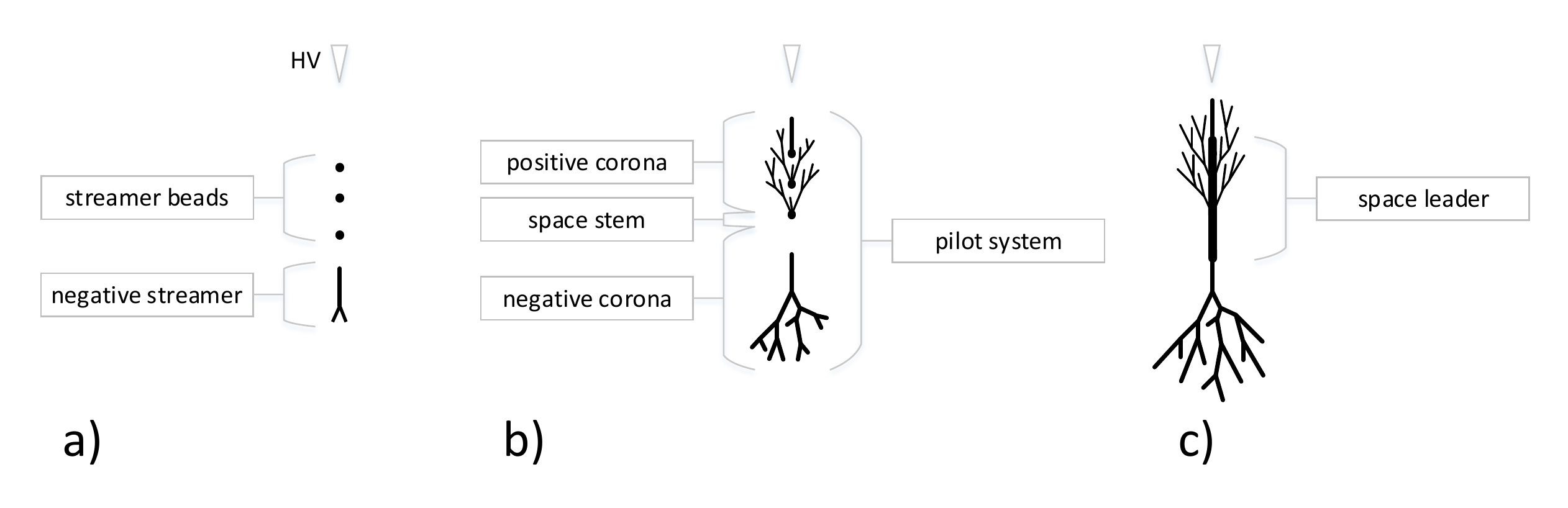}
\caption{Sketch of the pilot system development. The space stem on panel b) will appear as a bright spot with divergent streamers on streak photographs. In longer gaps such systems may develop into hot space leaders, as shown in panel c).}
\label{fig:sketch}
\end{figure*}

Figure~\ref{fig:sketch} summarizes the pilot system development. It starts with a negative streamer that leaves one or more beads behind, that may grow into $\Psi$s, and that finally transforms into the bipolar structure. An attempt to combine this history with the streak photographs of \cite{LesRenardieres1981} is hindered by the space-time convolution inherent to streak images. To this adds the larger gaps of 2 and 7~m versus ours of 1 up to 1.5~m and the longer voltage rise time of  6~$\mu$s versus 1.2~$\mu$s. Figure 6.1.5(b) in \cite{LesRenardieres1981} shows at least three stationary beads at 10 times larger separation than ours that last for about 0.5~$\mu$s; the beads seem to appear 'out of the blue' in virgin air. Our images demonstrate that a precursor streamer initiates the beads.

Though it is not apparent from the photographs presented here, previous observations on longer spark gaps show that the pilot system becomes a hot space leader if given sufficient time \cite{LesRenardieres1981}.
Figure 6.3.2 in \cite{LesRenardieres1981} shows that the instantaneous velocity of the space stem ranges from $2\times10^4$ to $2\times10^5$~m/s; the higher velocity goes with the shorter voltage rise time. Such space stem behavior can be explained by subsequent launching of positive streamers starting from the first bead to the last. Naturally, this will appear as a moving space stem on streak photographs.

\subsection{A detailed view of the cathode-directed part of the pilot}
\label{sec:positive_part}

\begin{figure*}[ht]
\centering
\includegraphics[height=8cm]{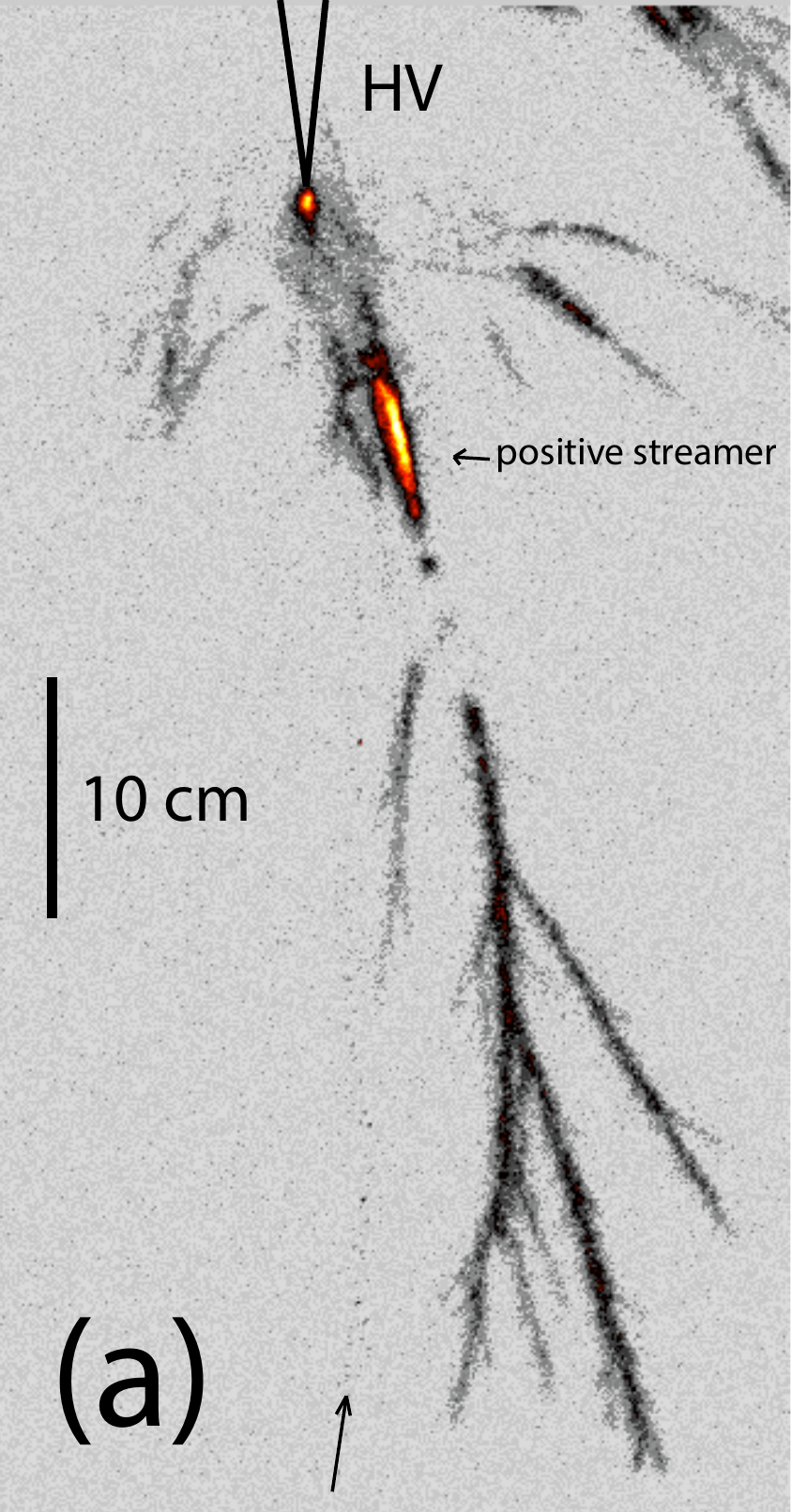}
\includegraphics[height=8cm]{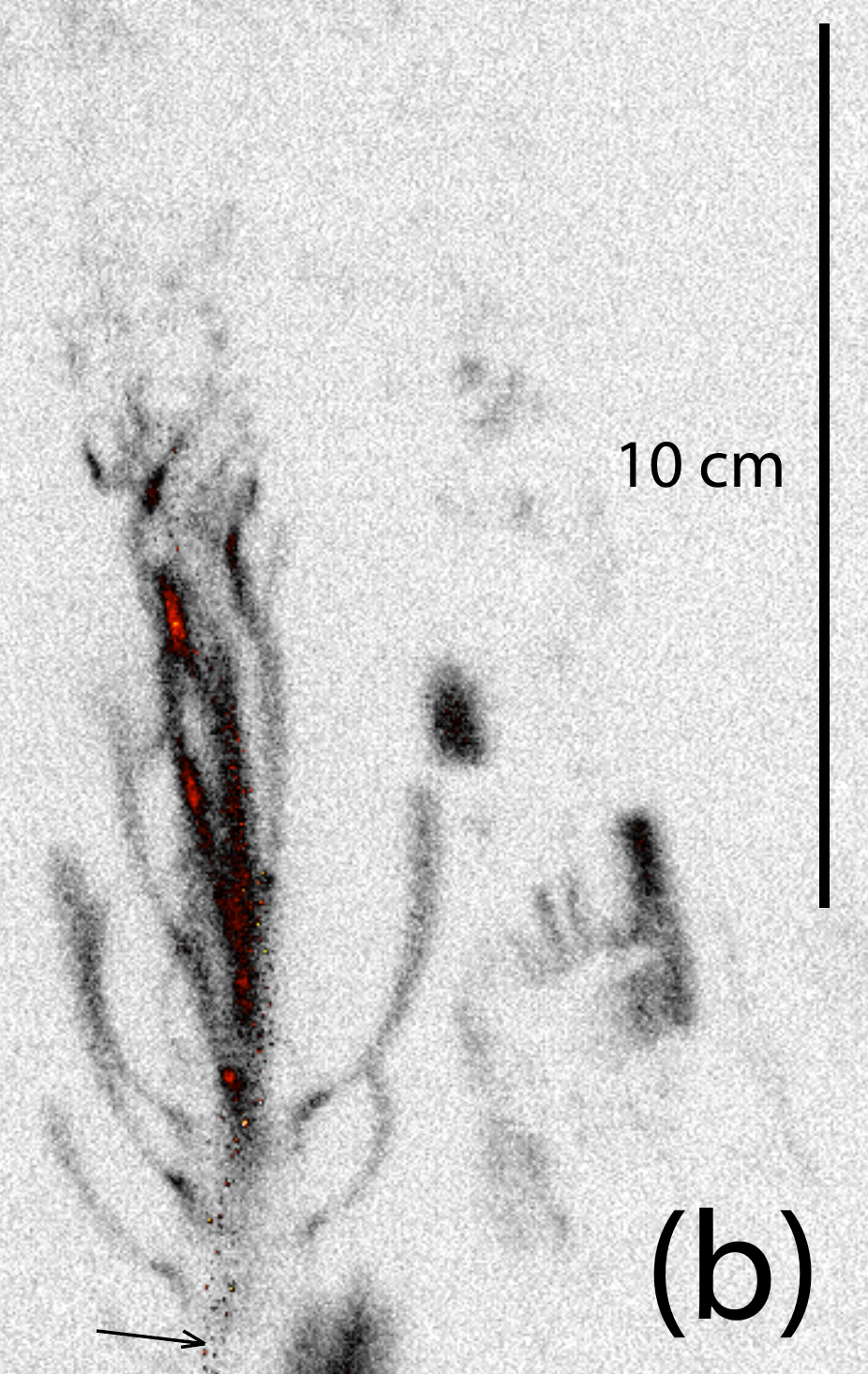}
\caption{(a) The top part of the pilot system contains only one positive streamer approaching the HV electrode. Exposure time is 100~ns. (b) Only the top part of a pilot system with many positive streamers originating from one column. Exposure time is 20~ns.}
\label{fig:positive_part}
\end{figure*}

Two types of positive corona emanate from the beads. The first is a single positive streamer, for instance image(a) in figure~\ref{fig:positive_part}; the second, our stack of $\Psi$s in image (b), has many streamers. The single positive streamer is significantly brighter and wider than all other streamers. The $\Psi$s  are shown in detail in figure~\ref{fig:positive_part} image (b), and also in figure~\ref{fig:collage} images (c) - (f). The streamers follow the local electric field lines towards the HV electrode. There are no visible beads anymore, which indicates that they fade away quickly, in fact about 10 times quicker than those in figure 6.1.5(b) of \cite{LesRenardieres1981}.

The faint speckle trace is visible in images (a) and (b) and indicated by small arrows. In image (a) it runs from the HV electrode tip down in the middle of the picture. In image (b) it comes from the left upper corner and goes through the structure down. It is a camera artifact. The camera's electronic shutter is switched off during the final breakdown, but some light can still leak through it and appear in the images as a trace. It helps to identify the streamer that grows into the final spark.\\

\subsection{Reconnection}

\begin{figure*}[ht]
\centering
\includegraphics[width=0.7\linewidth]{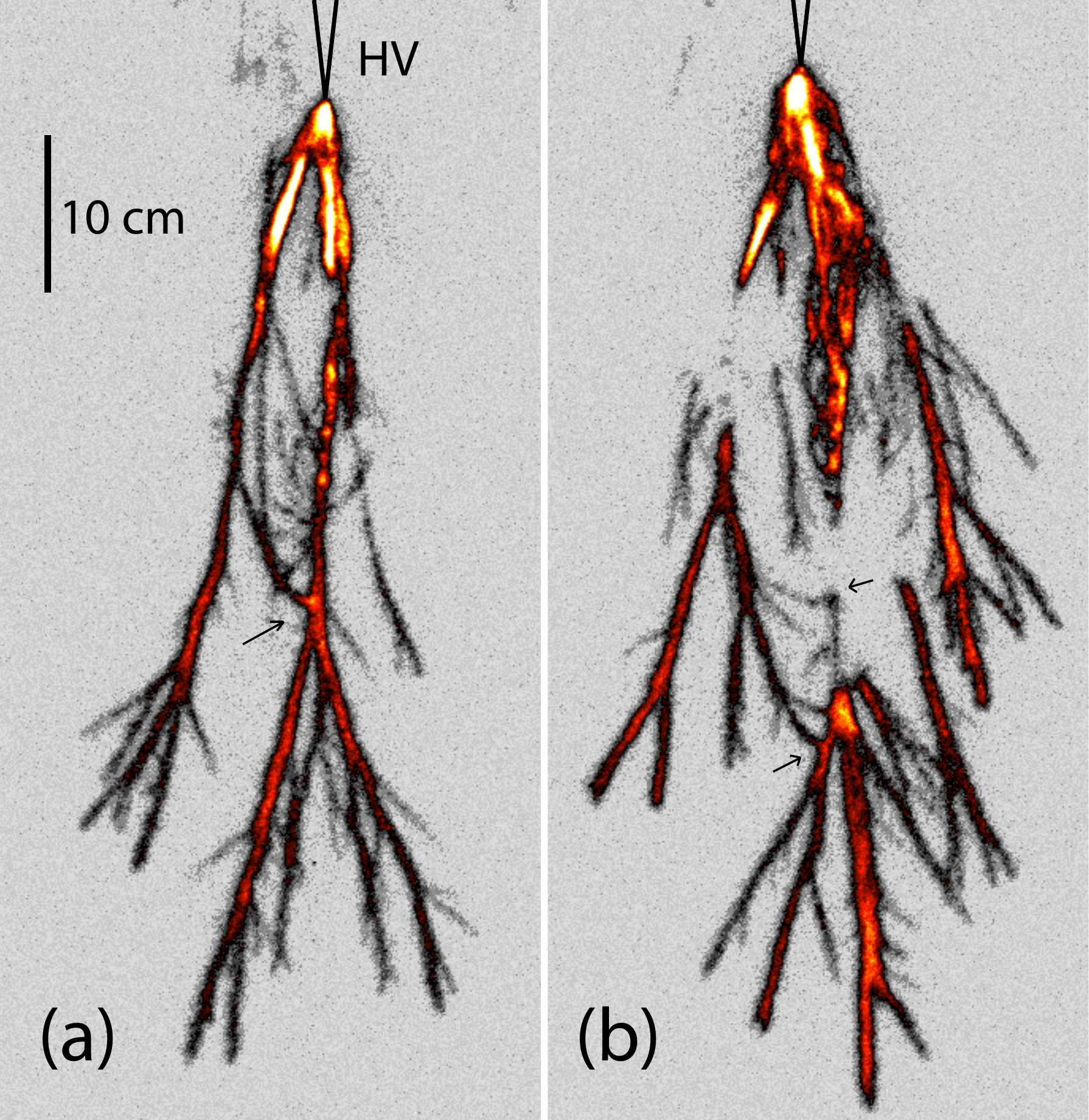}
\caption{The pilot system reconnections indicated by small arrows. Exposure time is 100~ns.}
\label{fig:reconnection}
\end{figure*}

Figure~\ref{fig:reconnection} shows an example of the pilot reconnection. Both types of pilot systems, as described above, are clearly visible.
A reconnection between positive streamers in STP ambient air was previously shown in \cite{Nijdam2009a} and possible mechanism proposed in \cite{Luque2014}. It is shown here that the reconnection also occurs between negative streamers, in this case interconnecting two pilot systems. The characteristic curvature of the streamer path and termination on the edge of another streamer channel indicates that they merged.

\section{1D model of the ionization evolution in the electrode gap}

\subsection{Model description}

\begin{figure}
\centering
\includegraphics[width=0.5\textwidth]{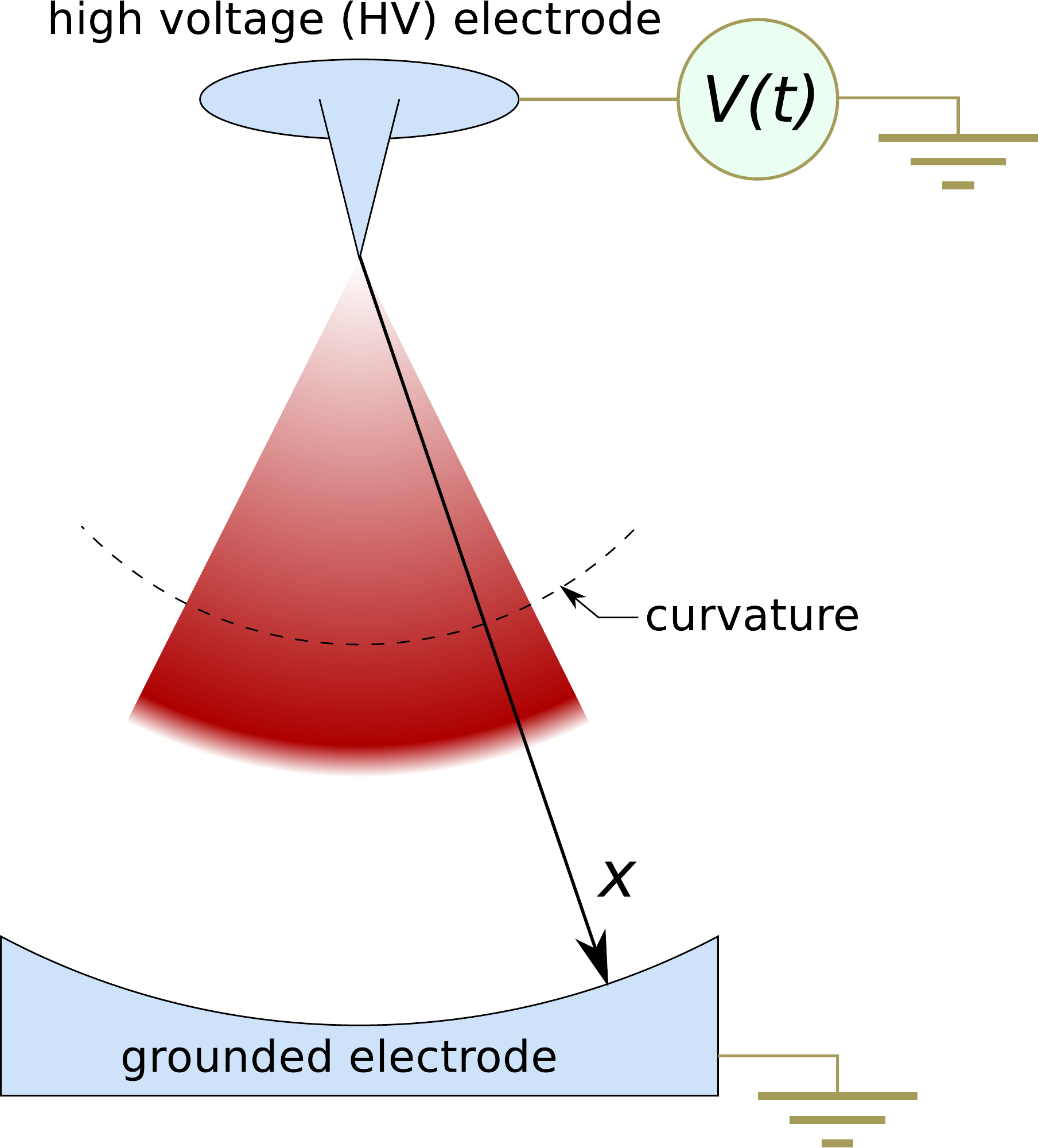}
\caption{Schematics of a 1D model of the laboratory spark discharge with effective curvature.}
\label{fig:model_cartoon}
\end{figure}

Numerical modeling of streamer development in air has currently reached a rather advanced stage. The streamers discharge has been modelled in full 3D space and based solely on microscopic physical mechanisms, e.g. \cite{Nijdam2014}. However, obtaining numerically in this way a fully developed branching fractal streamer pattern is computationally difficult and still under development.
To study the streamer structures with limited computational resources, one can introduce macroscopic physics, i.e., assumptions about the details of streamers which are not modelled microscopically \cite{Luque2014}. The ``pilots'' occur only at an advanced stage of a streamer discharge, by which we mean that the streamer corona have been fully developed and the individual streamers may have undergone possibly multiple branching. In order to understand the pilots, we are thus forced to follow the route of macroscopic (simplified) modeling. So that we can simplify the modeling, we make a rather crude assumption of spherical symmetry in the developed structured streamer discharge, with physical values being functions only of a single (radial) coordinate. The individual streamer branches are thus not considered but are treated collectively, and we represent the collective streamer effects in transverse (angular) direction in terms of the effective curvature, which is thus different from the curvature of individual streamer heads. The overall schematic representation of the model used is shown in figure~\ref{fig:model_cartoon}.

We shall model the development of a discharge in quasi-electrostatic approximation, i.e., neglecting the effects of electromagnetic waves. This can be done since the ratios of typical length and time scales, including  the typical streamer velocities $1.5\times10^6 - 4\times10^6$~m/s (given above in this paper), are much less than the speed of light. In principle, the relatively large $\sim1$~m size of the electrode gap can give importance to electromagnetic effects of very fast phenomena happening at time scales $\lesssim$1~ns, which could occur, e.g., during the streamer collisions; this is a topic for our future research. The quasi-electrostatic equations may be represented as:
\[
\left.
\begin{array}{rcl}
\nabla^2 \phi & = & - \rho/\varepsilon_0 \\
\dot{\rho} & = & -\nabla\cdot\left(\sigma\mathbf{E}\right) \\
\dot{N_e} & = & (\nu_i^\mathrm{eff}(E)-\nu_a^\mathrm{eff}(E))N_e+S_p
\end{array}\right\}
\]
In these equations, $\mathbf{E}=-\nabla\phi$, $\phi$, $\rho$ are electric field, potential, and charge density, respectively, $\sigma=e\mu_eN_e$ is the electric conductivity, $N_e$ is the electron density, $\nu_i^\mathrm{eff}$ and $\nu_a^\mathrm{eff}$ are effective ionization and attachment rates which describe propagation of streamers, which are different from the physical (microscopic) ionization and attachment rates (denoted here by $\nu_{i,a}$), and $S_p$ is the photoionization source.

This model includes a row of simplifying assumptions. First,  the electron mobility $\mu_e$ is assumed constant because it varies very little in a large range of electric fields, e.g., from $\sim 0.05$~m$^2$\,s$^{-1}$\,V$^{-1}$ at $E=0.3E_b$ to $\sim 0.04$~m$^2$\,s$^{-1}$\,V$^{-1}$ at $E=2E_b$,  where $E_b\approx3\times10^6$~V/m is the electric breakdown field \cite{Lisovskiy1998}.  We may therefore justify our assumption of constant electron mobility by arguing that the discharge-related processes occurring at $E\lesssim E_b$ are much less important than at fields $\sim E_b$. Second, we neglect ion conductivity due to the fact that ion mobility is at least by a factor of $\sim10^2$ smaller \cite{Horrak2000}. Third, we neglect electron advection effects, since the velocity of electron drift is much smaller than the streamer velocity. This, in particular, leads to having no difference in propagation of positive and negative ionization fronts in our model, while the observed velocities differed by a factor of $\sim$2 in the same background field. Fourth, we neglected electron diffusion $D_e$, valued at $\sim 0.08$--$0.16$~m$^2$/s in the range of electric fields of interest \cite{Morrow1997}, because the characteristic Kolmogorov-Petrovskii-Piskunov velocity of the ionization $2\sqrt{\nu_i(E_b)D_e}\sim10^4$~m/s \cite[p.~77]{Lagarkov1994} is also much smaller than the observed streamer velocities.

The electric breakdown occurs above $E_b$, where  the ionization prevails over attachment. However, in a 1D situation the propagation of negative streamers occurs above the negative streamer sustainment field $E_{-s}$, which is lower than $E_b$. Thus, we choose the functional dependence of the effective 1D rates $\nu^\mathrm{eff}_{i,a}(E)$ in such a way that the ionization occurs at $E>E_{-s}\approx 1.25\times 10^6$~V/m \cite{Moss2006}.  We approximate these rates with power functions:
\begin{equation}
\nu_i^\mathrm{eff}(E)=(E/E_{-s})^\alpha/\tau_i,\qquad \nu_a^\mathrm{eff}(E)=(E/E_{-s})^\beta/\tau_i
\label{eq:nu_powerlaw}
\end{equation}
where $\tau_i\approx 7$~ns is the typical ionization time.

We note that the empirical data for the physical values of $\nu_{i,a}$ \cite{Morrow1997} may also be fitted with power-law functions (\ref{eq:nu_powerlaw}), but we of course must use $E_b$ instead of $E_{-s}$ in these formulas. In the case of the physical values $\nu_{i,a}$, the best-fitting coefficients are $\alpha=5.5$ and $\beta=1.1$.

Another source of ionization growth is $S_p$, the photoionization, which is one of the mechanisms responsible for ionization front propagation (the other mechanisms being the neglected electron advection and electron diffusion). The photoionization is a non-local source
\[ S_p(\mathbf{r})=\int S_i(\mathbf{r}') F(\mathbf{r}-\mathbf{r}')\,d^3\mathbf{r}' \]
where $S_i=\nu_i^\mathrm{eff}(E)N_e$ is the ``local'' ionization rate.
We model it as the ``exponential profile'' model \cite{Luque2007}:
\[
F(r)=\frac{C}{\Lambda^2}\frac{e^{-r/\Lambda}}{4\pi r}
\]
where $\Lambda$ and $C\ll 1$ may be understood as the characteristic ``length'' and the ``strength'' of photoionization, respectively. This model is chosen for computational efficiency, because we can find $S_p$ as the solution of Helmholtz equation:
\[
(1-\Lambda^2\nabla^2)S_p=CS_i(\mathbf{r})
\]

The electrode gap is modelled as 1-dimensional interval $x\in[0,L]$, where the emitting electrode (the cathode in the case of the negative discharge study considered here) is at $x=0$, while the grounded electrode is at $x=L=1$~m. The system is assumed to be symmetric in the transverse direction. Voltage $\phi(0)=V(t)$ is applied at $x=0$, while $x=L$ is held at $\phi(L)=0$. The discharge is started by small initial ionization at $t=0$, $x=0$.

The curvature of ionization front $\varkappa$ is included through the expression for divergence of vectors $\parallel x$:
\[ \nabla\cdot\equiv \partial_x+\varkappa \]
For example, a spherically-symmetric system may be modelled by taking $\varkappa=2/r$. For the results presented here, we take a constant value $\varkappa=\mathit{const}>0$ for simplicity, which is equivalent to the transverse area of the discharge growing exponentially with distance. This may be justified by the streamers branching repeatedly with a fixed interval, and the transverse area being proportional to the number of streamers.

\subsection{Model results}
\label{ssec:model_results}

\begin{figure}
\includegraphics[width=\textwidth]{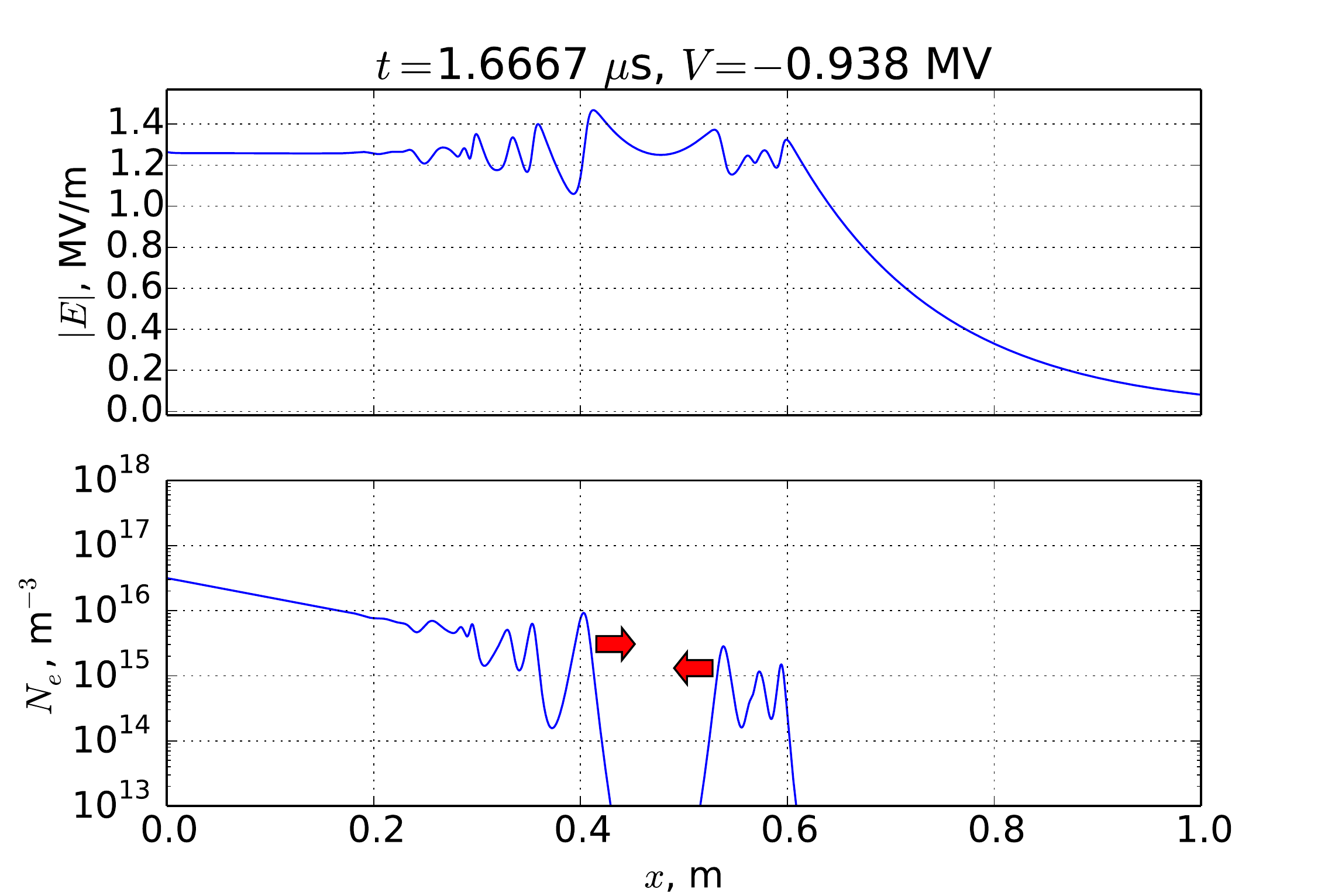}
\caption{Simulation results for  parameters: $V(t)=-(625+200[t/\mbox{$\mu$s}])$~kV, $\Lambda=0.005$~m, $C=0.01$, $\varkappa=7\mbox{ m}^{-1}=\mathit{const}$, $\alpha=3$, $\beta=1$. Arrows show the apparent movement of ionization enhancements.}
\label{fig:reverse_streamers}
\end{figure}

In the process of the nonlinear development of ionization growth, the ionization developed into multiple persistent peaks (seen in Figure~\ref{fig:reverse_streamers}) with complex dynamics (i.e., moving in both directions) which may be interpreted as various luminous features of streamers. In particular, the first peak in an ionization wave may be interpreted as the streamer head, while the consequent peaks, which move in the same direction or become stationary, may be interpreted as beads. Under some conditions (specified in the next paragraph) we observed a particular class of such peaks, which exhibited the following stages of evolution: (1)~an ionization peak  appears at the most advanced point of the ionization wave; (2)~the ionization is extinguished in the part of the gap between this peak  and the emitting electrode; (3)~as the voltage at the emitting electrode increases further, a reverse ionization wave separates from the peak and moves towards the emitting electrode, while a direct ionization waves moves towards it, until they collide; (4)~after even further voltage increase, the ionization peak continued its movement toward the grounded electrode. In view of the observations reported above, these ionization peaks  may be interpreted as the pilots which exhibit similar behavior, namely that they launch ionization waves in both directions: positive streamers towards the emitting electrode (cathode) and the negative streamers towards the grounded electrode (anode). Figure~\ref{fig:reverse_streamers} presents a snapshot of the process described above (at stage 3). Note that we used a slower-growing shape of ionization function ($\alpha=3$ instead of 5.5). The higher values of power coefficient $\alpha$ create steeper ionization edges. The simulations with other values of $\alpha$ were also performed (e.g., $\alpha=1$, $\beta=0$) and they also produce similar results (i.e., formation of the ``reverse streamers'' and the HV electrode current pulsations, see below). The chosen lower value of $\alpha$ in a 1D case creates a smoother front, and thus simulates the uncertainty in the position of the individual streamer heads (they may be distributed around some average position in $x$); unfortunately, the exact value of $\alpha$ which is best suited for this is not known to us at the current stage of research. We also chose the value of the photoionization length $\Lambda$ so that the simulation produces approximately the observed streamer velocities, while the value of $C$ is about the same as can be obtained from experimental data \cite{Penney1970}.

As we see, the variations in the ionization rate functional dependencies on $E$ did not change qualitatively the result of having ``islands'' of ionization and reverse ionization waves (stages 2--3 described above). By varying other parameters we preliminary conclude that this stages only appear when (1)~the voltage has a stage when it gradually increases with time and (2)~the photoionization effect is rather large. Although the propagation of streamers does need a large photoionization, and the ``islands'' of stage~2 appeared even at small values of $C$, the reverse ionization waves (stage~3) only appeared when $C$ exceeded a certain threshold.

\begin{figure}
\includegraphics[width=\textwidth]{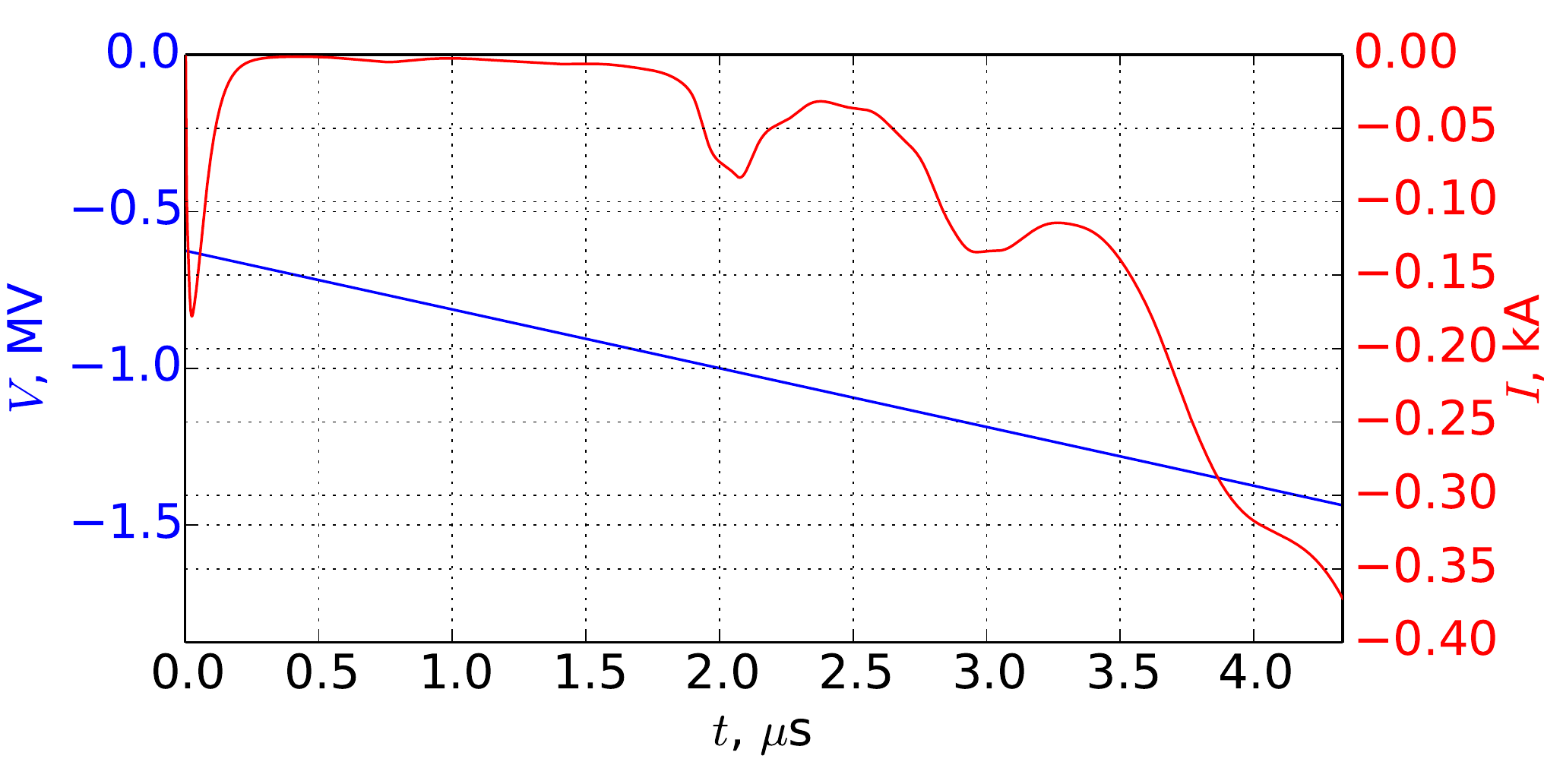}
\caption{The high-voltage electrode current. Simulation parameters are the same as in Figure~\ref{fig:reverse_streamers}, except $L=2$~m was taken in order to capture more pulses.}
\label{fig:current_pulses}
\end{figure}

Beside pilots and reverse streamers, another interesting outcome of the presented 1D model were the quasi-periodic current pulses at the high-voltage electrode (cathode), shown in Figure~\ref{fig:current_pulses}, which reproduce, at least qualitatively, the experimentally observed pulses shown in Figure~\ref{fig:plot}. The current was calculated taking into account both conductivity and displacement currents:
\[ I=A_0(\varepsilon_0\dot{E}+\sigma E)|_{x=0} \]
where the area of the cathode $A_0=0.03$~m$^2$ is chosen so that the current is of the same order as those in Figure~\ref{fig:plot}.

In the past, the pilots and stepping mechanism were studied numerically in long negative sparks (leaders) by 1D modeling \cite{Gallimberti2002,Rakotonandrasana2008}. However, these models treated separately the stages of the discharge which lead to the development of the pilots and stepping and therefore also take into account transition from the streamer corona to a leader discharge. In contrast to these models, we demonstrated that the leader development is not necessary for the pilot formation, and the similar stepping stages appear automatically from the nonlinear nature of the system without artificially subdividing the process. The colliding individual streamers were considered by \cite{Cooray2009} in the context of X-ray production by high electric field, but the streamer configuration was taken as an input on the basis of previous results and only $E$ field was calculated. The modeling of a streamer collision, taking into account the microscopic physical processes, was also performed by \cite{Ihaddadene2015}. We emphasize that, in contrast to the presented model, the last two works describe collision of two pre-existing streamers and do not deal with the development of the global streamer discharge system.

\section{Discussion}

\subsection{Monotonous vs bead-like structure of a streamer channel}

In most existing theoretical treatments (see chapter 12.3 in \cite{Raizer1991}) the streamer channel is understood as a linear structure along which the ionization (and therefore the electric field and current) vary monotonously. For example, the ionization is highest at the head of the streamer channel and gradually decreases in the backward direction. However, the high-speed videos of sprites \cite{Stenbaek-Nielsen2008} reveal luminous ``beads'' which move along each individual streamer channel. We may therefore propose a hypothesis that a streamer channel is not monotonous but has peaks of enhanced conductivity and/or field which manifest themselves as ``beads''. The presented simulation results supports this hypothesis. Namely, there are multi-peak structures in Figure~\ref{fig:reverse_streamers} even when the peaks are moving in the same direction (so that they are not parts of disconnected streamers).  However, another interpretation of multi-peak structures in the simulation results is also possible. Namely, since we modelled a whole group of streamers, the peaks may be just heads of separate monotonous streamer channels or groups thereof. The beads and periodic structures in ionization waves were attributed to the attachment instability by \cite{Luque2016}, who presented a linear-wave analysis of this phenomenon in their Appendix~A. The mechanisms included in the two models are slightly different (e.g., electron advection in \cite{Luque2016} vs. the effective curvature in the present work), so this similarity requires further investigation.

\subsection{Pilots-like structures in nature}

Pilots strikingly resemble high-altitude discharges known as sprites \cite{Stenbaek-Nielsen2008,Cummer2006}. Such peculiar features as glowing beads, streamer branching on beads, counter-propagating streamers originated from the beads, difference in brightness of the top and bottom parts and finally reconnection, characterise both phenomena. For visual comparison we refer to Figure~16 in \cite{Kochkin2014}. Two known types of sprites, carrot and column, can be directly compared to two types of pilots, as shown in section \ref{sec:positive_part}. Similarity is clear, despite the fact that in the laboratory pilots are pointed towards the sharp electrode tip, while sprites in nature originate from a dispersed charge region and appear vertically in 2D images.

The last concern of scientific community regarding different polarity of pilots and sprites has recently been eliminated. It has been known since 1999 that sprites are not uniquely associated with positive cloud-to-ground ($+$CG) lightnings, but can also be triggered by negative $-$CG flashes \cite{Barrington-Leigh1999}. However, the community continued to be s{\color{red}k}eptical in accepting sprite polarity asymmetry and existence of ``negative sprites'' [private conversations at AGU/EGU meetings]. Five more evidences of sprites following a negative $-$CG discharge were published in 2016 \cite{Boggs2015}. With simple considerations put forward in the next subsection, we cannot yet explain the existence of pilots with opposite polarity. In this case, new experimental studies of positive laboratory discharges with different voltage rise times are highly desirable.

\subsection{Role of pilots in the polarity asymmetry of leader propagation}

As it was mentioned in the Introduction, the pilots may transform into space leaders \cite{Gallimberti2002}. We may also speculate how pilots determine the polarity asymmetry between the modes of propagation of negative and positive leaders. Let us consider the conditions for formation of a system of forward and reverse streamers moving towards each other, such as one that appeared during stage~3 of the simulation described in section~\ref{ssec:model_results}, and consider the differences for two different leader polarities.

In a negative leader discharge, the electrostatic field in front of the leader converges towards its tip. The negative forward streamers, being closer to the electrode, experience a higher field than the positive reverse streamers that are initiated from a position further away from the leader tip and are on their way to encounter the negative forward streamers. This is consistent with the fact that negative streamers need a higher field to support their propagation than the positive streamers. Thus, both forward and reverse streamers may exist at the same time.

On the other hand, in a positive leader discharge, the forward streamers are positive while the reverse streamers (if they appeared) would be negative. We see that the lower field, which the reverse negative streamers would experience, cannot support their propagation. Thus, we do not expect formation of reverse streamers in the positive-leader case, the ionization gap (if it appeared) would be filled only by forward positive streamers.

The role of the decreasing field in the differences between positive and negative leader propagation was also discussed by \cite{Gallimberti2002}, where they suggested that in the negative leader corona the forward-moving electrons attach in the lower field, which interrupts the current and leads to stepping.

\subsection{Role of pilots in X-ray production}
\label{ssec:discuss_xray}
Another implication of experiments and modelling is that the positive and negative streamers, which travel towards each other, will collide at a certain moment of the discharge. This may lead to a increased electric field in the gap between them, which subsequently leads to generation of high-energy electrons \cite{Østgaard2016}. It has been proposed as the possible mechanism of X-ray production in laboratory spark discharges \cite{Cooray2009}.

This mechanism has been modelled by \cite{Ihaddadene2015} who found that the number of X-ray photons produced may not be sufficient to explain observations. However, we think that the electric field in \cite{Ihaddadene2015} has been underestimated. The boundary conditions with fixed potential, represent a perfectly conducting boundary and lead to image charges, whose field reduces the field in the modelling domain. Thus, the actual field may be higher and also occupy a bigger volume. On the other hand, a higher field between the two colliding streamers would also lead to an increased streamer velocity, which would reduce the time of the existence of the region with high field. Since there are two counteracting effects, it is not clear without repeating the full simulation whether corrected boundary conditions would lead to increase of X-ray production. Here, we may note that the observations confirmed the coincidence of occurrence of X-rays in laboratory sparks with colliding positive and negative streamers \cite{Østgaard2016}, even if the exact mechanism of electron acceleration is still open for discussion.

\section{Conclusions}

In this work we first tracked a single pilot system development in the laboratory between two conical electrodes under 1~MV applied voltage. It was demonstrated for the first time that pilots do not develop from ``nowhere'', as was thought before \cite{Bazelyan2000,LesRenardieres1981}, but from isolated streamer beads, created in the wake of the negative streamer head. The beads, in principle, can be called a ``space stem'' for consistency with the previous studies, but it is important to highlight that they do not appear in virgin air, but behind a negative streamer.

The 1D model of the ionization front evolution demonstrated that such beads and reverse ionization waves can appear with certain photoionization parameters. However, we have not yet demonstrated the differences in the discharge polarity, as the electron drift was neglected; this is a subject of a future work.

Taking into account many similarities between pilot systems and sprites, not only in appearance but also in progression, we conclude that these are two manifestations of the same phenomenon. It is very desirable to investigate the pilot system development under different conditions, i.e. temperature, pressure, voltage rise time etc. In addition, the main modelling parameters as electron and photon density, their energy spectrum, the local E-field and potential with respect to ground remain hidden in all measurements. Some specially designed probes would solve some ambiguities.

\section*{Acknowledgement}

This work was supported by the European Research Council under the European Union's Seventh Framework Programme (FP7/2007-2013)/ERC grant agreement n.~320839 and the Research Council of Norway under contracts 208028/F50 and 223252/F50 (CoE). Pavlo Kochkin acknowledges financial support by STW-project 10757, where Stichting Technische Wetenschappen (STW) is part of the Netherlands organization for Scientific Research NWO.

\clearpage

\providecommand{\newblock}{}

\bibliographystyle{iopart-num}

\end{document}